\documentstyle[12pt,openbib,psfig]{article}
\newcommand{\al}{\alpha}

\newcommand{\de}{\delta}

\newcommand{\la}{\lambda}
\newcommand{\om}{\omega}

\newcommand{\si}{\sigma}

\newcommand{\La}{\Lambda}
\newcommand{\Om}{\Omega}

\newcommand{\f}{\frac}

\def\be{\begin{equation}}
\def\ee{\end{equation}}

\begin{document}

\title{\bf Decoupling of pion coupling $f_{\pi}$ from quarks at
high density in three models, and its possible observational
consequences.}

\author{ Manjari Bagchi $^{1,~*}$, Monika Sinha
$^{1,~\dagger}$, Mira Dey $^{1,~*}$ and   Jishnu Dey $^{2,~*,~
\ddagger}$}

\maketitle

\begin{abstract}

{  Chiral symmetry is restored at high density, quarks become
nearly massless and pion, the Goldstone of the symmetry breaking
decouples from the quarks. What happens at high density is
important for finding the density dependence of Strange Quark
Matter (SQM), - which in turn is relevant for understanding the
structure of compact stars.}

\end{abstract}

\vskip .5cm


\vskip .5cm

\noindent $^1$ Dept. of Physics, Presidency College, 86/1 ,
College Street, Kolkata 700 073 and IUCAA, Pune, 411007, India.\\
$^2$ UGC Research Professor, Dept. of Physics, Maulana Azad
College, 8  Rafi Ahmed Kidwai Road, Kolkata 700 013 and IUCAA,
Pune, 411007, India.\\ $^*$ Work supported in part by DST grant
no. SP/S2/K-03/2001, Govt. of India.\\
 $\dagger$ CSIR Research fellow, Govt. of India. \\ $\ddagger$ Project : Changing Interface of
Nuclear, Particle and Astrophysics,  email : kamal1@vsnl.com.
\newpage
\section {Introduction}

~~~We investigate what happens to $f_\pi$, at high densities. In
our convention, $f_\pi$ is defined as follows (vacuum value of
$f_{\pi} ~\sim~ 93$ MeV) :

\be <0|A^a_\mu(x)|\pi^b(q)>~=~iq_\mu \de_{ab} f_\pi(q^2)e^{-iqx}.
\label{fpi} \ee

The Hellmann-Feynman theorem, applied to a nuclear many body
model, gives the quark condensate in nuclear matter at high
density \cite{plb}. Coupling this with theoretical one of Nambu
and Jona-Lasinio (NJL) \cite{njl}, one can extract $f_\pi(n_B)$,
where $n_B$ is the baryon number density. Chiral Symmetry breaking
and pion properties was discussed in the framework of NJL model by
Bernard \cite{ber}. There was follow up of the work on
$f_\pi(n_B)$, using the NJL model, by Bernard, Meissner and Zahed
\cite{bermeis} and more recently by Caldas \cite{caldas}.

Low temperature QCD sumrule results also give $f_\pi(n_B)$ upto
$n_B~\sim~ 4~n_0 $ where $n_0$ is normal density \cite{dd}.

Again density dependent quark masses, used for SQM calculations
\cite{D98}, can be used to fix the parameters of the NJL model.
This in turn enables one to get the pion coupling to the QCD
vacuum $f_\pi(n_B)$.

The quark mass is given in the SQM \cite{D98} as :

\be M_i^* = m_i + M_Q ~~sech\left( \f{n_B}{Nn _0}\right), \;\;~~~
i = u, d, s. \label{qm} \ee

\noindent where $n_B = (n _u+n _d+n _s)/3$ is the baryon number
density; $n _0 = 0.17~fm^{-3}$ is the normal nuclear matter
density; $n _u$, $n _d$, $n _s$ are number densities of u, d and s
quarks respectively and $N$ is a parameter. The current quark
masses ($m_i$) are taken as : $m_u = 4 \;MeV,\; m_d = 7 \;MeV,\;
m_s = 150 \;MeV$. $M_Q$ is the constituent quark mass taken around
$\sim ~ 325$ MeV according to latest version of the model
\cite{SubMnj}.

\section{Nuclear matter model}

In the relativistic $\si-\omega$ models of nucleon matter it is
found that the quark condensate can be estimated using the
Hellmann-Feynman theorem and this was investigated in detail in
\cite{dm,plb}. Interestingly, the title of \cite{plb} also
referred to a decoupling, - that of the nucleon mass and the quark
condensate in the medium. The Walecka model, the pioneering one,
implies an effective quark condensate that increases with density.
This is contrary to common belief. The newer Zimanyi-Moskowski
(ZM) model, has an edge over the Walecka model in satisfying the
criterion that the quark condensate falls with increase in density
as shown in \cite{plb}.

Further relevance of the ZM has been recently pointed out by Sinha
et al., who have shown that the velocity and the incompressibility
of the ZM model also match onto a quark model \cite{plbnew}.

According to Hellmann-Feynman theorem \cite{marz,cohentannaou,dm}

\be
<\psi(\la)\mid\f{d}{d\la}H(\la)\mid\psi(\la)>~=~\f{d}{d\la}<\psi(\la)\mid
H(\la)\mid\psi(\la)>, \label{hf} \ee where $H(\la)$ is any
hermitian operator depends on a real parameter $\la$ and
$|\psi(\la)>$ is a normalized eigen vector of $H(\la)$.

In QCD the Hamiltonian density is given by

\be {\cal H}_{QCD}~=~{\cal H}_0~+~2m_q\bar qq, \ee with the major
part being the chirally symmetric ${\cal H}_0$. Here $m_q$ is
quark mass and $q$ is the quark field.

Making the identification $H~\rightarrow~\int d^3x~{\cal H}_{QCD}$
and $\la~\rightarrow~m_q$ one finds the Hellmann-Feynman theorem
as:

\be 2m_q~<\psi(\la)\mid\int d^3x~\bar qq\mid\psi(\la)>~=~m_q\f{d}
{dm_q}<\psi(\la)\mid\int d^3x~{\cal H}_{QCD}\mid\psi(\la)>.
\label{hfqcd} \ee

The above equation may be applied to nuclear matter and vacuum
with $\mid\psi(\la)>~=~\mid n_B>$ and $\mid\psi(\la)>~=~\mid vac>$
respectively. Here $\mid n_B>$ denotes ground state of nuclear
matter at rest with nucleon density $n_B$ and $\mid vac>$ denotes
the vacuum state. taking the difference of the above two cases and
keeping in mind the uniformity of the system, one gets

\be 2m_q~(<\bar qq>_{n_B} ~-~<\bar qq>_{vac})~=~m_q\f{d{\cal
E}}{dm_q} \label{nm}. \ee where $n_B$ is the number density in
nuclear matter.

Here in general $<\Om>_{n_B}~=~<n_B\mid\Om\mid n_B>$ and
$<\Om>_{vac}~=~<vac\mid\Om\mid vac>$ notations have been used for
an arbitrary operator $\Om$.

The energy density ${\cal E}$ of nuclear matter is given by

\be {\cal E}~=~ n_B M_N ~+~\de {\cal E}, \label{enden} \ee where
$\de {\cal E}$ is the contribution to energy density from the
nucleon kinetic energy and nucleon-nucleon interaction energy.
$\de {\cal E}$ is of higher order in the nucleon density and is
empirically small at low density.

At low density the quark condensate can be related to the nucleon
$\si$ term $\si_N$ which may be defined as \cite{jaffe}

\be \si_N~=~\f13~\sum_{a=1}^3(<N\mid[Q_A^a,[Q_A^a,H_{QCD}]]\mid
N>-<vac\mid[Q_A^a,[Q_A^a,H_{QCD}]]\mid vac>), \ee where $Q_A^a$ is
axial charge, $H_{QCD}$ QCD Hamiltonian and $\mid N>$ is state
vector of nucleon at rest. Alternatively $\si_N$ can be expressed
as:

\be \si_N~=~2m_q\int d^3x~(<N\mid \bar qq \mid N>-<vac\mid \bar qq
\mid vac>) \label{sigmandef}. \ee

\noindent where

\be \si_N~=~m_q\f{dM_N}{dm_q} \label{sigman}. \ee

\noindent Hence Eq. (\ref{nm}) can be written as (using Eq.
(\ref{enden}))

\be 2m_q~(<\bar qq>_{n_B} ~-~<\bar qq>_{vac})~=~m_q~n_B
\f{d{M_N}}{dm_q}+m_q\f{d~\de \cal E}{dm_q} ~=~n_B \si_N
+m_q\f{d~\de \cal E}{dm_q} \label{8}. \ee

\noindent Assuming translational invariance which makes quark
condensate constant one can define \be \si_A~=~2m_qV~(<\bar qq
>_{n_B}-<\bar qq>_{vac}) \label{sigmanm}. \ee

\noindent Using Eq. (\ref{8})

\begin{eqnarray} \si_{eff}~=~\f{\si_A}{A}~=~
\si_N\left(1+\f{d~\de ({\cal E} / {n_B})}{dM_N}\right)
\end{eqnarray}

Now from Gell-Mann$-$Oakes$-$Renner relation we know

\be 2m_q~<\bar qq>_{vac}~=~-m_\pi^2f_\pi^2 \label{gor} \ee

\noindent $m_\pi$ and $f_\pi$ being the pion mass and pion decay
constant respectively. From Eq. (\ref{nm})

\begin{eqnarray}
& &
  \f{<\bar qq>_{n_B}}{<\bar
qq>_{vac}}~=~1-n_B \f{\si_{eff}}{m_\pi^2f_\pi^2}
\end{eqnarray}

In the ZM model the Lagrangian describes the motion of a baryon
with an effective mass instead of bare mass. This information goes
to modify the scalar coupling constant making it density dependent
while the vector coupling remains the same. In contrast with the
Walecka model ${<\bar qq>_{n_B}}/{<\bar qq>_{vac}}$ goes down with
density \cite{plb}.

\section{The QCD sumrule method}

This is a very elegant method devised by Shifman, Vainshtein and
Zakharov \cite{SVZ} and consists of equating the coupling of an
interpolating Lorentz invariant current for a meson or a baryon -
with proper spin, parity and isospin degrees of freedom - to
quark-antiquark for meson and three quark for baryons. The quarks
or antiquarks are then allowed to mix into the QCD vacuum - which
have condensates of quark-antiquark and gluons - and also exchange
gluon lines  through operator product expansion (OPE). Starting at
high momentum transfer for finding the coefficients of the OPE by
Borel transform one finds a `window' where the sum rule becomes
independent of he Borel mass parameter. The condensate values
picked up from one set, say the $\rho$ meson can be used for all
the meson or baryon sumrules. For meson baryon coupling one has to
go over to three point functions which is more complicated but
straight forward in principle. Reviews are available by Reinders,
Rubinstein and Yazaki \cite{rry} and Dey and Dey \cite{book}.

Working out the density and temperature dependence of the
pion-nucleon coupling constant ($g_{\pi NN}$), within the
framework of QCDSR techniques, Dey and Dey  \cite{dd} deduced the
$f_{\pi}$ to be about half its value (44 MeV compared to 81 MeV)
at four times normal density {\footnote { $f_{\pi}$ was normalized
to vacuum value 130 MeV in \cite{dd} and is readjusted here by the
factor $\sqrt 2$. It is somewhat low in a nucleon which already
has a substantial hadron density. }}. This agrees with the
estimate of the present paper using the NJL model. The sumrule
model predicted that the Goldberger-Treiman relation $g_{\pi
NN}~=~ M_N\sqrt{2}/f_\pi$ is independent of density \cite{sr} and
this was confirmed  in a later calculation by \cite{EletskyKagan}.

\section{Quark mass used in stellar calculation}

Early suggestion of a cosmic separation of phases of hadronic and
strange matter led to investigations properties of strange quark
star, but were not very successful. This was because the star with
maximum mass had a radius of about $\sim 9~-~ 10$ kms and this is
comparable to that of a neutron star. One could not distinguish
between the two. The density dependence of quark masses was not
considered in these early models. At high density there is chiral
symmetry restoration (CSR) and the masses approach the current
quark mass values.

By putting CSR, in a simple tree level large $N_c$ model
\cite{D98}, one can set up an equation of state (EOS) and seek to
explain the properties of compact stars Her X$-$1 and 4U
1820$-$30. Li and others used this EOS to explain   the properties
of SAX J1808.4$-$3658 or 4U 1728$-$34 \cite{Li99a, Li99b}. Compact
stars are assumed to be composed of (u, d, s) matter that is very
dense (typically 4.6 (surface) to 15 (core) times the normal
nuclear density). In the model (u,d) matter has less binding per
baryon E/A, compared to $Fe^{56}$ and (u, d, s) matter has more.

It is interesting to note that many X-ray emitters are rotating
and shows periodicity. Only recently, however, six sources were
discovered starting with SAX J1808.4$-$3658 (1998), which are
accreting millisecond X-ray pulsars and an important question is
raised by Wijnands \cite{rudy} : why are those compact stars
different from others for which no pulsations have been found?
Perhaps, he comments, new ideas need to be explored to explain
these six sources. Stability of the star may be a crucial point in
resolving this issue, according to the present authors  and the
use of (u, d, s) matter with restored chiral symmetry may help. We
must mention that the model leads to stars which are very stable
as shown by Sharma et al. \cite{sharma}, by matching the external
Schwarzchild metric to a realistic one at the boundary of the
star. For details we refer the reader to \cite{sharma}. Strange
star models, with the above EOS, are also very stable when
rotating fast, as shown by Gondek-Rosi\'nska et al. \cite{gondek}
and Bombaci, Thampan and Datta \cite{btd}. The density dependence
of the strong coupling constant $\al_s$ in this model was explored
using the simple Schwinger-Dyson expansion advocated by Bailin,
Cleymans and Scadron in Ray et al. \cite{rdd}.

Further, there are other interesting applications of this model
enumerated below :

\begin{enumerate}

\item  X-ray superbursts lasting for several hours thrice in 4U
1636$-$53 and once (so far) in KS 1731$-$260 \cite{mnras},  and
also the phenomenon in general, seen in 7 stars altogether.

\item Occurrence of two quasi periodic peaks in the X-ray power
spectrum model of 4U 1636$-$53 and  KS 1731$-$260 \cite{Li99b,
bani} and other stars.

\item Absorption in 1E1207.4$-$5209 \cite{mpla} and emission
\cite{mnras2} in various stars like 4U 0614$+$091, 2S 0918$-$549,
4U 1543$-$624, 4U 1850$-$087 from surface compressional modes.

\end{enumerate}

In addition the interesting model of quark nova of Ouyed et al.
\cite{o}, employs the idea of contraction of normal matter when it
is converted to (u, d, s) matter of the above model. Gravitational
energy from matter falling onto a compact core, formed during a
supernova explosion and consequent generation of a core remnant,
can lead to gamma ray after glow according to \cite{o}.

The density dependent quark mass used in the (u, d, s) matter is
used to generate the pion coupling to quarks in the present paper.

\section{The Nambu Jona-Lasinio model}

~~~We recall that in the model of Nambu and Jona-Lasinio, one can
calculate  the quark mass $M^*$, $f_\pi$, the quark condensate
$<\bar qq>$ for a given coupling $G$, -  following the equations
below in terms of a cut off $\La$ of 631 MeV (see \cite{plb3}) :

 \be
M^*~=~m_0~+~4G(N_c~N_f+\f12)M^*\int^\La \f{d^3p}{(2\pi)^3}\f1E
\label{mass} \ee

\be f^2_\pi ~=~N_c~M^{*2}\int^\La \f{d^3p}{(2\pi)^3}\f1{E^3}
\label{fpi2} \ee

\be <\bar qq>~=~ <\bar uu>~=~ <\bar dd>~=~ -6M^*\int^\La
\f{d^3p}{(2\pi)^3}\f1E \label{quark} \ee

Knowing the NJL coupling G, one can therefore relate the
quantities $M^*,~f_{\pi}$ and $\bar qq$.  We assume that $G$
varies with density and find it (1) by fitting it to $f_{\pi}$ in
the QCDSR for which we do not need the NJL model, (2) using $<\bar
qq>$ in the $\si-\omega$ nuclear matter model and (3) by fitting
density dependent (u,d,s) quark mass in equation (\ref{mass}).
>From (\ref{fpi2}) and density dependence of G, $f_\pi$ and the
corresponding quark condensates are obtained and are tabulated in
table \ref{gcon}.

\begin{table}[htbp]

\caption{Variation of $f_\pi$, $G$,  and $<\bar qq>$ with density
ratio ($n_0~ = ~0.17/fm^3$).}

\vskip .5cm

\begin{center}

\begin{tabular}{|c|c|c|c|}
\hline${n_B}/{n _0}$ &$f_{\pi}$&G&$<\bar q q>^{1/3}$\\
&$(MeV)$&$(MeV^{-2})$&$(MeV)$\\ \hline
       1 &    90.9227 &$4.936\times10^{-6}$&$-243 $ \\ \hline
       2 &   85.8209& $4.682\times10^{-6}$&$-232$\\ \hline
       3 &  77.5264&$ 4.389\times10^{-6}$&$-223 $\\ \hline
       4 &  67.1299 &$ 4.13\times10^{-6}$&$-208 $\\ \hline
       5 &  56.1471 & $ 3.922\times10^{-6}$&$-191 $\\ \hline
       6 &   45.8124 &$3.755\times10^{-6}$&$-174 $\\ \hline
       7 &   36.8022&$3.606\times10^{-6}$&$-159 $\\ \hline
       8 &  29.3312&$ 3.458\times10^{-6}$ &$-144  $\\ \hline
       9 &  23.3414&$ 3.294\times10^{-6}$&$-131 $\\ \hline
      10 &  18.6502&$ 3.103\times10^{-6}$&$-120 $\\ \hline
      11 &  15.0386&$ 2.877\times10^{-6}$&$-110 $\\ \hline
      12 &  12.2944 &$ 2.616\times10^{-6}$&$-102$\\ \hline
      13 &  10.2314&$ 2.323\times10^{-6}$&$-95 $\\ \hline
      14 &  8.69422&$ 2.01\times10^{-6}$&$-89 $\\ \hline
      15 &  7.55736&$ 1.692\times10^{-6}$&$-84 $\\ \hline
      16 &  6.72196 &$ 1.386\times10^{-6}$&$-80 $\\ \hline
      17 &  6.11141 &$ 1.107\times10^{-6}$&$-78 $\\ \hline
      18 &   5.66718&$8.645\times10^{-7}$&$-75$\\ \hline

\end{tabular}

\end{center}

\label{gcon}

\end{table}

At high density,  nucleon mass decreases very much with $f_{\pi}$
in the Skyrme and other models and the nuclear radius becomes so
large that there is no point in talking of a `confined' nucleons,
the quarks are percolating.

\begin{figure}[htbp]
\centerline{\psfig{figure=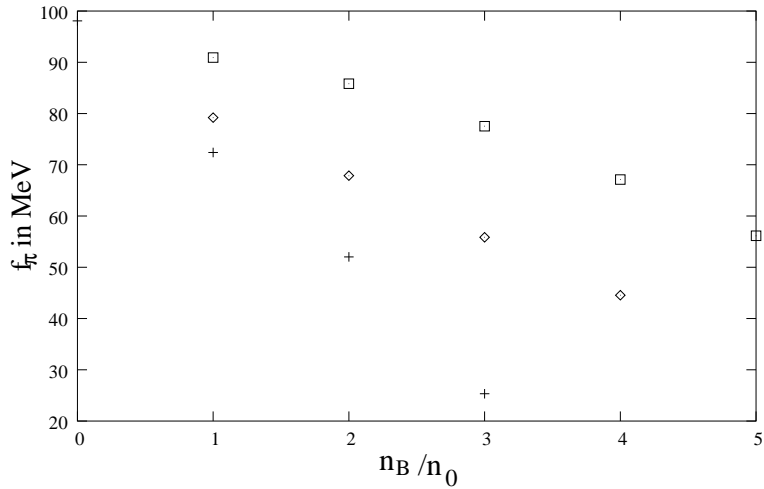,width=10cm}}
\caption{\footnotesize{Density  $f_{\pi}$ from different models
upto ($\sim 4 \rho$) : diamonds corresponds to QCDSR results, $+$
corresponds to the nuclear matter model of ZM, squares correspond
to the SQM.}}
 \label{njlcomp}
\end{figure}

\begin{figure}[htbp]
\centerline{\psfig{figure=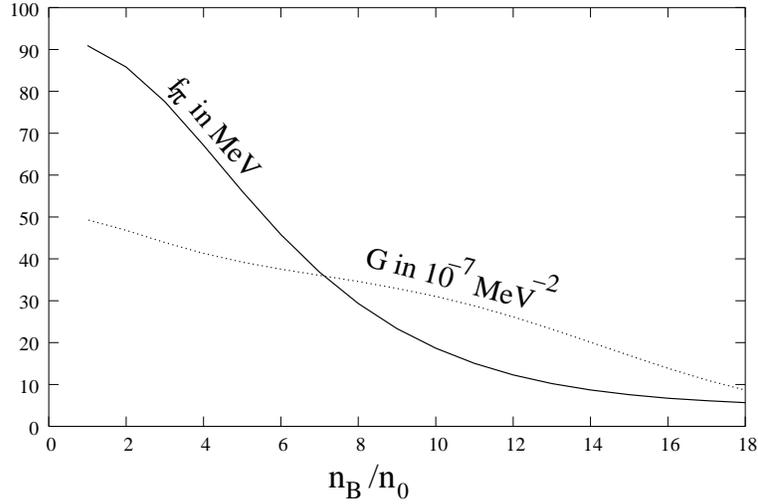,width=10cm}}
\caption{\footnotesize{Predictions from density dependent quark
mass of the SQM upto high density : density ($\rho$) dependence of
$f_{\pi}$ (full line), $\rho$ dependence of the NJL coupling
constant G (dotted line ). }}
 \label{njlfull}
\end{figure}

Fig. (\ref{njlcomp}) shows that the $\si-\om$ model predicts a
zero $f_\pi$ at about $\sim~4\rho_0$. The QCDSR fall-off is also
sharp compared to SQM. We can thus claim that the CR in SQM is
mild. The full $n_B$ dependence is shown in Fig. (\ref{njlfull})
where the density dependence of NJL coupling G is also shown. Our
result checks with \cite{bermeis}. For example, for number density
five times $n_0$ the value of $f_\pi$ is about 60 MeV. A much more
mild density dependence of $f_\pi$ is implied by Caldas
\cite{caldas} who display a number like 80 MeV. It will be very
interesting to see if the photon width increase, predicted in this
paper, is indeed found in heavy ion collisions. The photon
momentum resolution of STAR experiment does not allow any decisive
conclusion about the possible enhancement of the $\pi^0$ width, -
for details see \cite{caldas}.

For future use we have fitted all the quantities by the equation
\be y~=~ a_1x+a_2x^2+a_3x^3+a_4x^4+a_5x^5+a_6x^6+a_7x^7+a_8x^8
\label{genfit} \ee

where y represents the variables ($~f_\pi,~G,~<\bar qq>$
respectively ) and x is the density ratio ${n_B}/{n_0}$. The
coefficients for each quantity are tabulated in table \ref{gcoef}.

\begin{table}[htbp]

\caption{Coefficients for density expansion of $f_\pi$, $G$ and
$<\bar qq>$.}

\vskip .5cm

\begin{center}

\begin{tabular}{|c|c|c|c|}
\hline
coef&$f_{\pi}$&G&$<\bar q q>$\\

\hline $a_1$&164.51 &$9.025\times10^{-6}$&$-2.65893\times10^{7}$ \\
\hline

$a_2$&-104.31&$-5.832\times10^{-6}$&$ 1.72333\times10^{7}$ \\
\hline

$a_3$& 31.08 &$1.780\times10^{-6}$ &$-5.11645\times10^{6}$ \\
\hline

$a_4$&-5.21 &$-3.006\times10^{-7}$ & 849758.0   \\
\hline

$a_5$& 0.513 &$2.959\times10^{-8}$ &-83014.7    \\
\hline

$a_6$&-0.029 &$-1.690\times10^{-9}$&  4724.5   \\
\hline

$a_7$&-0.0009  &$5.187\times10^{-11}$ & -144.734 \\
\hline

$a_8$&$-1.156\times10^{-5}$&$-6.611\times10^{-13}$&1.843 \\
\hline

\end{tabular}

\end{center}

\label{gcoef}

\end{table}

\section{The problem of relativistic heavy ion collisions (RHIC)}

Recently exciting new results have been reported by several groups
from the gold on gold nuclear collisions in Brookhaven. It appears
that there is thermalization and a high temperature is reached.
The problem with the experimental results is that although the
system is not describable by hadronic models, the nearly
non-interacting quark gluon model also does not seem to work. In
the language of the protagonists ` the interpretation of current
data relies heavily on theoretical input and modeling, in
particular on the apparent necessity to include partonic degrees
of freedom in order to arrive at a consistent description of many
of the phenomenon observed in experimental data. Seen from a
purely experimental point of view this situation is somewhat
unsatisfying, but probably not unexpected, not avoidable,
considering the complexity of the reaction and associated
processes \cite{brahms}.' Quoting another group to conclude `the
data from RHIC collisions provide strong evidence for the creation
of high energy density, low baryonic chemical potential, medium
which can not simply be described in terms of hadrons and whose
constituents experience significant interactions with each other
\cite{phobos}.'

In conclusion, from high temperature RHIC data, it is not clear
that either of  the features of QCD like chiral symmetry
restoration (CSR) or asymptotic freedom (AF) is actually realized
due to the complexity of the system and the system may display
strong interacting coherent partonic interactions. The system that
one can observe in stars may in fact yield a clearer signature of
CSR and AF. We are grateful to the referee for allowing us to
comment on this feature.

In the next section we shall discuss the nature of the density
dependence that one expects from heuristic considerations given by
various authors.

 \section{Discussion}

In the model \cite{bl}, the radius of the pion is :

\be R_ \pi= 0.4 \sqrt{z}~/f_{\pi}\ee

\noindent where z is the probability of finding a purely $\bar q
q$ component in the pion. The decrease of $f_{\pi}$ with
increasing density signifies increase in the radii of the hadrons.
This in fact ultimately leads to the percolation of the quarks.
Assuming the nucleon radius $R_N ~=~ c~MeV~ fm~/f_{\pi}$, in
\cite{dd} the constant $c$ is adjusted to get the radius of the
nucleon at normal density :
 \be R_N~=~86.12~MeV~fm/f_{\pi} \label{rn}. \ee

\noindent One can review QCD scales following Bailin, Cleymans and
Scadron \cite{BCS}.

\be m_{dyn}~\simeq ~ \La_{\bar{MS}} e^{\f16}~\simeq ~ 300 MeV
\label{bcs} \ee

\noindent where the minimal subtraction overall energy scale of
QCD, $M_{\bar{MS}}\simeq~250$ MeV for the 3 flavour case. This is
close to 325 MeV of (\ref{qm}). One can go on to get \be f_\pi~=~
\f{\sqrt3}{2\pi} m_{dyn}~\approx ~ 87 MeV \ee and the string
tension \be \si ~\approx ~\sqrt{\f\pi2} m_{dyn} ~ \approx ~400
MeV. \ee

As density increases $f_\pi$ and quark condensate decreases with
$m_{dyn}$ and this is borne out by the NJL model of the present
paper.

 $f_{\pi}$ is a parameter in chiral models, the pioneering one
being the Skyrme model,

\be {\cal
L}~=~\f{1}{8}f_{\pi}^2~Tr(\partial_{\mu}U\partial_{\mu}{U^{\dag}})
+(1/32e^2)~Tr(\partial_{\mu}U{U^{\dag}},\partial_{\nu}U{U^{\dag}})^2
\ee

\noindent where it is the {\bf only} parameter depending on
temperature and density \cite{ddg}. The consequences of
$f_{\pi}(\rho)$ was first analyzed by Rho \cite{rho} and Meissner
\cite{meiss}.

The nucleon  mass can be calculated using the Skyrme model  :

 \be M_N~=~f_{\pi}{\mu}/e\sqrt{2} \label{rn2} \ee

\noindent where $\mu~=~73.0$  is an integral over the chiral angle
of the Skyrmion \cite{anw} and $e$ is the dimensionless Skyrme
parameter taken to be 5.78.

In this context it is interesting to emphasize the suggestion by
Dosch and Narison \cite{dosch}, from QCD Sum Rule (QCDSR) method,
that $e$ is independent of the quark condensate. Based on this
\cite{ddg} found that indeed the parameter $e$, being independent
of temperature and density, could in fact be 2$\pi$, as suggested
by Skyrme to represent a spin current.  Incidentally the nucleon
radius $R_N$ is proportional to its inverse $R_N ~=~ c~MeV~
fm~/f_{\pi}$ where c is a constant. In general,  all chiral model
properties scale with $f_{\pi}$, as in the Skyrme model.

These values have the support of tentative observations made for
compact stars. The importance of the results can be anticipated,
since a convincing proof for the existence of such compact stars
may soon emerge, from the copious flow of recent astrophysical
observations.

In particular it will be interesting to see if there is any change
in  Ouyed's model for Skyrmion Star \cite{ouyed} with a density
dependent $f_{\pi}$.

\section{Conclusions and summary}

~~~We have calculated the variation of the pion coupling
$f_\pi(\rho)$ with density in the Nambu Jona-Lasinio model and it
is satisfying to see that this matches with expectations of other
models. $f_\pi(\rho)$ and the constant $G$ are parametrized as
polynomials of  density in the hope that the results may be used
in future calculations.

In summary,  we have calculated the pion coupling constant
$f_{\pi}$ from the density dependent (u, d, s) masses employed in
compact star models and the results are qualitatively matching
with other models namely (1) QCD sumrule and (2) nuclear matter
models. Results may be useful for chiral models where use of
$f_{\pi}(\rho)$ will produce significant difference at high
density.

To conclude, in our opinion, observations on high density matter -
perhaps possible in compact stars in an indirect manner -  may
yield signatures of asymptotically free and nearly chirally
symmetric matter. These signatures are elusive in present day RHIC
data.





\begin{thebibliography}{}


\bibitem{plb} A. Delfino, J. Dey, M. Dey, M. Malheiro, Phys. Lett. B363, (1995) 17.

\bibitem{njl} Y. Nambu and G Jona-Lasinio, Phys. Rev. 122 (1961)
345.

\bibitem{ber} V. Bernard, Phys. Rev D34 (1986) 1601.

\bibitem{bermeis} V. Bernard, U. Meissner and I. Zahed, Phys. Rev D36 (1987) 819.

\bibitem{caldas} H. Caldas, Phys. Rev C69 (2004) 035204.

\bibitem{dd} J. Dey and M. Dey, Phys. Lett. B 176, (1986) 469.




\bibitem{D98} M. Dey, I. Bombaci, J. Dey, S. Ray  and B. C.
Samanta, Phys.  Lett. B438, (1998) 123; Addendum   B447, (1999)
352; Erratum B467, (1999) 303;  Indian J. Phys. 73B, (1999) 377.

\bibitem{SubMnj}``Strange star equation of state with modified Richardson
potential.", M. Bagchi, S.Ray, J. Dey and M. Dey (to be
published.)

\bibitem{dm} T. D. Cohen, R. J. Furnstahl and D. K. Greigel, Phys. Rev. C45, (1992) 1881.

\bibitem{plbnew} M. Sinha, M. Bagchi, J. Dey, M. Dey, S. Ray and S. Bhowmick,
Phys. Lett. B590, (2004) 120; hep-ph/ 0212024.

\bibitem{marz} E. Marzbacher, {\it Quantum Mechanics} (Wiley,
Newyork, 1970) 2nd Ed.

\bibitem{cohentannaou} C. Cohen-Tannoudji, B. Diu and F.
Lalo\"{e}, {\it Quantum Mechanics} (Wiley, Newyork, 1977) Vol. II.

\bibitem{jaffe} R. L. Jaffe and C. L. Korpa, Commun. Nucl. Part. Phys. {\bf17}, 163
(1987).



\bibitem{SVZ}M. A. Shiffman, A. I. Vainshtein and V. I. Zakharov,
Nucl. Phys. B. 147 (1979) 385, 448, 519.

\bibitem{rry}L. J. Reinders, H. R. Rubenstein and S. yazaki, Phys.
Rep. C. 127 (1985) 1.

\bibitem{book} M. Dey and J. Dey, ``Nuclear and Paricle Physics
The Changing Interface " , Springer - Verlag (1993), Chapter 9.

\bibitem{sr} J. Dey, M. Dey and P. Ghose, Phys. Lett. B 166, (1985) 181.

\bibitem{EletskyKagan} V. L. Eletsky and I. I. Kogan, Phys. Rev. D49, (1994)
3083.


\bibitem{Li99a} X. Li, I. Bombaci, M. Dey, J. Dey  and E. P. J. van den
Heuvel, Phys. Rev. Lett. 83, (1999) 3776.


\bibitem{Li99b} X. Li, S. Ray, J. Dey, M. Dey  and I.  Bombaci,
Astrophys. J. 527, (1999) L51.

\bibitem{rudy} R. Wijnands, `` Accretion-driven X-ray millisecond pulsars ", astro-ph/0501264.

\bibitem{sharma} R. Sharma, S. Mukherjee, M. Dey and  J. Dey, Mod. Phys. Lett. A 17, (2002) 827.

\bibitem{gondek} D. Gondek-Rosi\'nska, T. Bulik, L. Zdunik, E.
Gourgoulhon, S. Ray, J, Dey and M. Dey, Astron. \& Astrophys.
 363, (2000) 1005; astro-ph/0007004.

\bibitem{btd} I. Bombaci, A. V. Thampan and B Datta, Ap. J. 541,
(2000) L71.

\bibitem{rdd} S. Ray, J. Dey, M. Dey, Mod. Phys. Lett. A15, (2000)
1301.

\bibitem{mnras} M. Sinha, M. Dey, S. Ray and J. Dey,  Mon. Not. Roy.
Astron. Soc. 337, (2002) 1368.

\bibitem{bani} B. Mukhopadhyay, S. Ray, J. Dey and M. Dey,
Astrophys. J. 584, (2003) L83.


\bibitem{mpla} M. Sinha, J. Dey, M. Dey, S. Ray and S. Bhowmick,
Mod. Phys. Lett. A18, (2003) 661.

\bibitem{mnras2} S. Ray, J. Dey, M. Dey and S. Bhowmick,
Mon. Not. Roy. Astron. Soc. 353, (2004) 825; astro-ph/0406162.

\bibitem{o} R. Ouyed, J. Dey and M. Dey, Quark Nova,
Astron. \& Astrophys. Lett. 390, (2002) L39-42;
astro-ph/0105109v3.


\bibitem{plb3} A. Delfino, J. Dey and M. Malheiro, Phys.Lett. B348, (1995)

\bibitem{brahms} I. Arsene et al.,  BRAHMS collaboration, nucl-ex/0410020 to be published in Nucl. Phys. A.


\bibitem{phobos} B. B. Back et al.,  PHOBOS collaboration, nucl-ex/0410022 to be published in Nucl. Phys. A.

\bibitem{bl} S. J. Brodsky and G. P. Lepage, Phys. Scr. 23, (1981)
945.

\bibitem{BCS} D. Bailin, Cleymans and Scadron, Phys. Rev. D31, (1985) 164.

\bibitem{ddg} J. Dey, M. Dey and P. Ghose, Phys. Lett. B193, (1987)
98.

\bibitem{rho} M. Rho, Phys. Rev. Lett 54  (1985) 767.

\bibitem{meiss} Ulf. G. Meissner, Nucl. Phys. A503, (1989) 801.


\bibitem{anw} G. Adkins, C. Nappi, E. Witten, Nucl. Phys. B228,
 (1983) 552.

\bibitem{dosch} H. G. Dosch and S. Narison, Phys. Lett. B180, (1986) 390.

\bibitem{ouyed} R. Ouyed, `` From the Skyrme model to hypothetical
Skyrmion stars: Astrophysical impiatons", astro-ph/0402122.





\end{thebibliography}
\end{document}